\documentclass{article}

\PassOptionsToPackage{numbers, compress}{natbib}
\usepackage[preprint]{neurips_2026}

\usepackage[utf8]{inputenc} % allow utf-8 input
\usepackage[T1]{fontenc}    % use 8-bit T1 fonts
\usepackage{hyperref}       % hyperlinks
\usepackage{url}            % simple URL typesetting
\usepackage{booktabs}       % professional-quality tables
\usepackage{amsfonts}       % blackboard math symbols
\usepackage{nicefrac}       % compact symbols for 1/2, etc.
\usepackage{microtype}      % microtypography
\usepackage{xcolor}         % colors
\usepackage{graphicx}
\usepackage{enumitem}
\usepackage{multirow}
\usepackage{colortbl}
\usepackage{pifont}

\usepackage{amsmath}
\definecolor{gold}{HTML}{FFDE78}    % 1st
\definecolor{silver}{HTML}{D9D9D9}  % 2nd
\definecolor{bronze}{HTML}{E6CCB2}  % 3rd
\newcommand\eg{{\it e.g.}}

\title{UniER: A Unified Benchmark for Item-level and Path-level Exercise Recommendation}

\author{%
  Xinghe Cheng$^\spadesuit$,\quad
  Guiyong Zhuang$^\spadesuit$,\quad
  Yusheng Xie$^\spadesuit$,\quad
  Jiapu Wang$^\heartsuit$,\quad
  Yixin Liu$^\diamondsuit$,\quad \\
  \textbf{Quanlong Guan}$^{\spadesuit*}$,\quad 
  \textbf{Liangda Fang}$^{\spadesuit*}$,\quad
  \textbf{Shirui Pan}$^\diamondsuit$\\
  $^\spadesuit$Jinan University, Guangzhou, China \\
  $^\heartsuit$Beijing University of Technology, Beijing, China \\
  $^\diamondsuit$Griffith University, Gold Coast, Australia \\
  \texttt{\{chengxh@stu2023, gql, fangld\}.jnu.edu.cn} \\
}

\begin{document}

\maketitle
{
\renewcommand{\thefootnote}{}
\footnotetext{\hspace{-0.6em}$^*$ Corresponding author.}
}

    \begin{abstract}
    Personalized exercise recommendation dynamically aligns pedagogical resources with individual knowledge mastery, which is crucial for satisfying students' dynamic learning needs in modern education.
    The field is currently driven by two dominant paradigms: Item-Level Exercise Recommendation (ILER) optimizes for immediate single-step state transitions, while Path-Level Exercise Recommendation (PLER) constructs coherent learning paths to maximize cumulative gains.
    Despite sharing the same ultimate objective, disparate evaluation setups have kept these two lines of research isolated, hindering unified benchmarking and fair comparison.
    To fill the gap, in this paper, we present a \underline{\textbf{Uni}}fied Benchmark for \underline{\textbf{E}}xercise \underline{\textbf{R}}ecommendation (UniER), a comprehensive evaluation framework that unifies ILER and PLER.
    Specifically, we introduce Weighted Cognitive Gain (WCG) as a unified metric to measure cross-paradigm algorithmic performance.
    Our benchmark encompasses 9 datasets spanning four generation methods, facilitating the comparison of 18 representative ILER/PLER methods.
    Through multi-dimensional analyses covering effectiveness, generalizability, robustness, and efficiency, our results reveal the systematic dominance of PLER and expose the pedagogical failure of ILER's fragmented recommendations under extreme sparsity and noise.
    Furthermore, we provide an open-source codebase of UniER to foster reproducible research and outline potential directions for future investigations\footnote{\href{https://github.com/chanllon/UniER}{https://github.com/chanllon/UniER}}.
    \end{abstract}

    \section{Introduction}
    \looseness=-1
    Exercise Recommendation (ER) has become a cornerstone of personalized learning, transforming how educational resources are aligned with individual learner needs~\cite{CheZF2025, LiuTL2019}. 
    The fundamental objective of ER is actively providing appropriate exercises to assist students in acquiring knowledge, thereby eliminating the need for self-directed searches.
    Although ER shares the notion of ``recommendation'', it comes with significant challenges that are not present in more conventional recommendation domains, such as e-commerce and media recommendation~\cite{RajJ2025, GreCG2024, ZhaCW2024, CheLR2023}. 
    In particular, interaction pattern mining techniques that are effective in conventional recommendations fail to reflect students' knowledge states, and optimizing solely for user preferences is insufficient to promote knowledge mastery.
    Consequently, effective ER requires paradigms that extend beyond behavioral similarity and preference optimization to directly support knowledge acquisition.
    
    \begin{figure}
        \centering
        \includegraphics[width=1\textwidth]{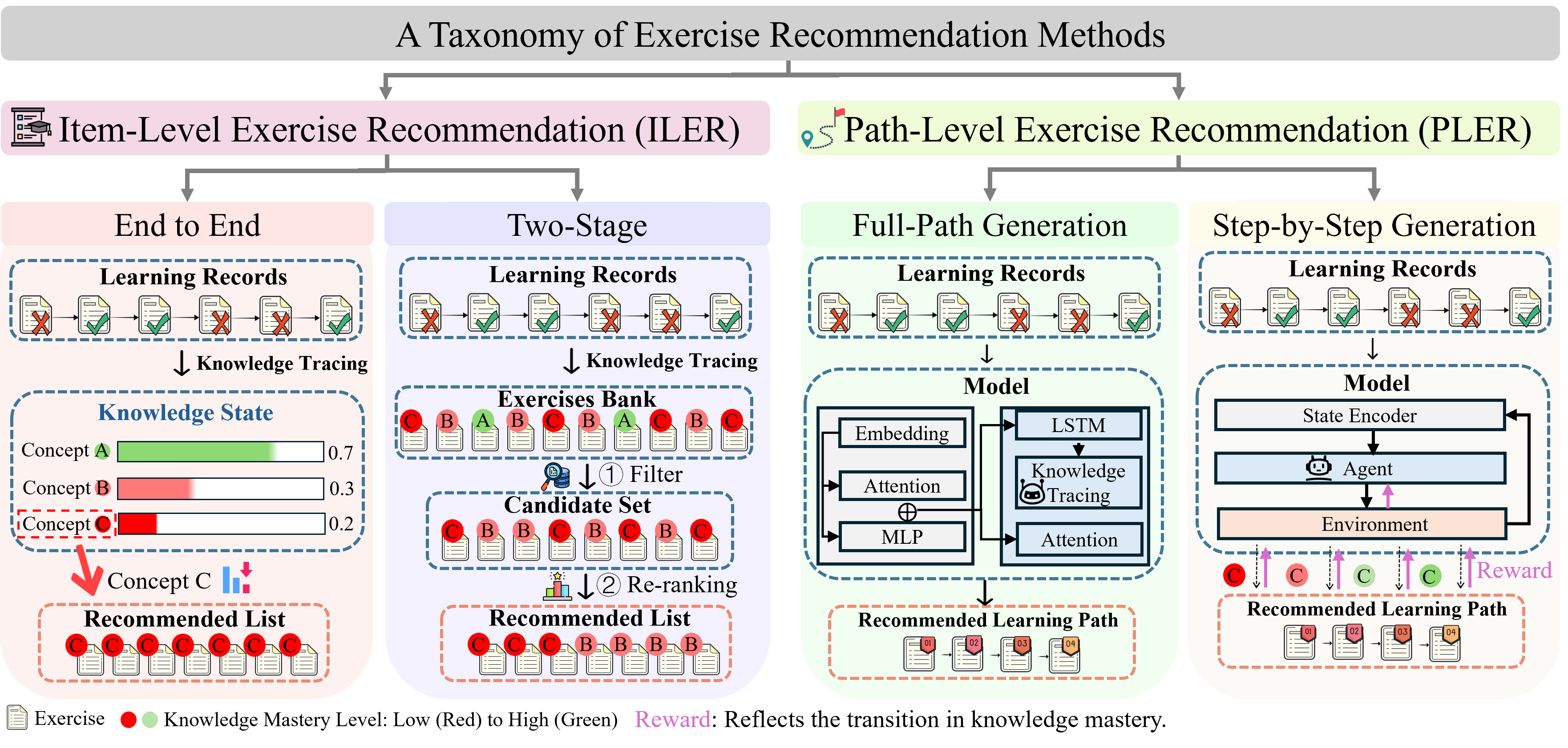}
        \caption{A Taxonomy of Exercise Recommendation Methods.}
        \label{exercise recommendation framework}
    \end{figure}

    \looseness=-1
    As one of the representative ER paradigms, Item-Level Exercise Recommendation (ILER) follows the traditional item-based recommender system paradigm, standing as a classical research direction~\cite{WuLT2020, GuaXC2023, CheZF2025}.
    Specifically, ILER formulates the task as a next-step recommendation process, aiming to retrieve an unordered set of exercises tailored to the student's immediate cognitive needs.
    Existing ILER methods can be categorized into two branches, namely end-to-end approaches and two-stage approaches. 
    End-to-end approaches seamlessly project unmastered knowledge concepts into the exercise space and make recommendations directly. 
    To further maximize the concept diversity within the final recommendation list, two-stage approaches explicitly decouple the process into filtering and re-ranking phases. 
    Nevertheless, existing ILER methods treat the recommendation process as isolated steps rather than a coherent learning path, which neglects the continuous nature of learning processes. 

    \looseness=-1
    Driven by the necessity to construct coherent learning paths, another line of research, Path-Level Exercise Recommendation (PLER), draws increasing attention in recent years~\cite{LiuTL2019, LiXY2024, CheZW2026}.
    PLER aims to generate an ordered sequence of exercises, meaning it is designed to maintain cognitive smoothness, and indicating a primary focus on achieving predefined pedagogical goals.
    Existing PLER approaches usually follow two paradigms: full-path generation and step-by-step generation. 
    Full-path generation methods employ an encoder-decoder framework to output the entire learning path in a single step by capturing latent dependencies between exercises.
    In contrast, step-by-step generation methods based on reinforcement learning (RL) model learning path generation as a Markov Decision Process (MDP)~\cite{PatST2021}, dynamically generating personalized paths aligned with students' cognitive development patterns.
    Compared to ILER, PLER places greater emphasis on modeling long-term sequential dependencies and goal-oriented learning planning. However, it often relies on predefined paths, thereby neglecting the pedagogical flexibility and immediate feedback inherent in individual learning processes.

    \looseness=-1
    Despite the methodological diversity in current ER research, the evaluation of ER approaches remains inadequate and incomplete, mainly due to two key challenges. 
    \textit{\textbf{Challenge 1}: {Lack of comprehensive evaluation protocols}.}
    As ILER and PLER paradigms diverge fundamentally in both predictive mechanisms and core objectives, they inherently rely on incompatible evaluation metrics.
    For instance, ILER predominantly relies on generic recommendation metrics that are fundamentally agnostic to pedagogical applicability.
    Even when incorporating KT, such evaluations prioritize isolated concept mastery, thereby neglecting the continuous dynamics of the learning process.
    In contrast, while PLER explicitly models the learning path, it lacks a comprehensive characterization of the overall knowledge framework, heavily relying on aggregated path-level rewards for evaluation.
    This disjointed evaluation protocol yields one-sided performance characterizations, hindering a comprehensive understanding of practical algorithmic efficacy.
    \textit{\textbf{Challenge 2}: Lack of unified horizontal comparisons}.
    Current studies report outcomes under disparate datasets and heterogeneous settings, lacking a unified benchmark to facilitate fair cross-method and cross-paradigm comparisons.
    Given the intrinsic differences between the ILER and PLER paradigms, direct comparison is inherently precluded.
    Consequently, although these methods have achieved remarkable progress within their specific settings, evaluating their relative strengths and applicable scenarios remains challenging, thereby creating a bottleneck for effective real-world deployment.
    
    \looseness=-1
    To fill the gap, in this paper, we develop a \textbf{Uni}fied Benchmark for \textbf{E}xercise \textbf{R}ecommendation, abbreviated as UniER.
    To overcome \textbf{\textit{Challenge 1}}, grounded in the Item Response Theory (IRT)~\cite{Lor1952, Lor2012}, we propose Weighted Cognitive Gain (WCG), a unified metric to quantify algorithmic performance for holistic ER evaluation. 
    WCG models cognitive gain through customizable weights, allowing us to evaluate each algorithm from multiple perspectives by instantiating different evaluation principles via varying weight configurations. 
    To address \textbf{\textit{Challenge 2}}, building upon WCG, we establish a comprehensive benchmark UniER based on 9 datasets to fairly compare 18 ILER and PLER methods under a unified experimental setting, exploring the performance landscape of state-of-the-art approaches across diverse scenarios and domains.
    In terms of \textbf{effectiveness}, we conduct a cross-paradigm empirical study within a unified benchmark, systematically uncovering the performance trade-offs between item-level accuracy and path-level coherence.
    To assess \textbf{generalizability}, we stress-test existing models under varying degrees of data sparsity and cold-start conditions, thereby revealing the inherent limitations of current paradigms under data scarcity.
    Furthermore, we examine model \textbf{robustness} by introducing stochastic noise into educational datasets, which allows us to quantify the resilience of these methods against data perturbations.
    Finally, we benchmark \textbf{efficiency} by quantifying training/inference latency and peak memory footprint, providing insights into the practical scalability of representative ER approaches.
    At the end of this paper, we discuss the future directions of this emerging research.
    
    \noindent\textbf{Key findings.} Through comprehensive cross-paradigm benchmarking, we summarize four remarkable observations: 1) PLER systematically outperforms ILER by maximizing long-term cumulative gains; 2) PLER structurally resists extreme noise via inherent pedagogical logic, exposing ILER's fragility; 3) Streamlined RL-driven methods exhibit unexpectedly high efficiency, drastically outperforming heavy-architecture baselines; and 4) The true computational bottleneck stems from coupling expensive graph searches with policy optimization, not the RL paradigm itself.

    \section{Preliminaries}
    \looseness=-1
    In this section, we formally define two core paradigms of exercise recommendation (ER), namely Item-Level Exercise Recommendation (ILER) and Path-Level Exercise Recommendation (PLER).
    
    \looseness=-1
    \noindent\textbf{Unified Problem Formulation.}
    The primary objective of ER is to recommend a subset of suitable exercises based on a student's historical interaction sequence.
    Let $\mathcal{S} = \{s_1, s_2, \dots, s_{|\mathcal{S}|}\}$ denote the set of students, $\mathcal{C} = \{c_1, c_2, \dots, c_{|\mathcal{C}|}\}$ denote the set of Knowledge Concepts (KCs), and $\mathcal{E} = \{e_1, e_2, \dots, e_{|\mathcal{E}|}\}$ denote the set of exercises, where $|\mathcal{S}|$, $|\mathcal{C}|$, and $|\mathcal{E}|$ represent their respective total numbers.
    Furthermore, the relationship between exercises and KCs is represented by a binary Q-matrix $\mathbf{Q} \in \{0, 1\}^{|\mathcal{E}| \times |\mathcal{C}|}$, where $Q_{i,j} = 1$ indicates that exercise $e_i$ involves KC $c_j$, and $Q_{i,j} = 0$ otherwise.
    The historical interaction sequence of student $s \in \mathcal{S}$ up to time step $t$ is defined as $\mathcal{H}_{s,t} = \{(e^{(1)}, r^{(1)}), (e^{(2)}, r^{(2)}), \dots, (e^{(t)}, r^{(t)})\}$, where $r^{(i)} = 1$ indicates that the student correctly answered exercise $e^{(i)} \in \mathcal{E}$, and $r^{(i)} = 0$ otherwise.
    
    \looseness=-1
    \noindent\textbf{Item-Level Exercise Recommendation (ILER).}
    The ILER paradigm primarily aims to recommend exercises that target a student's cognitive weaknesses.
    To achieve this, conventional approaches leverage Knowledge Tracing (KT) techniques~\cite{PieBH2015, MaYW2025} to diagnose unmastered KCs and retrieve candidate exercises covering these concepts \cite{HuaLZ2019, GhoHS2020, LiuLC2023, LiuHL2023}.
    Specifically, the KT model estimates the student's mastery over KCs from the historical interaction sequence $\mathcal{H}_{s,t}$, resulting in a continuous knowledge state vector $\mathcal{M}_{\text{KT}}(\mathcal{H}_{s,t}) \in [0, 1]^{|\mathcal{C}|}$.
    The KC-level knowledge gaps are represented as $(\mathbf{1} - \mathcal{M}_{\text{KT}}(\mathcal{H}_{s,t}))$ and mapped to the exercise space via the Q-matrix $\mathbf{Q}$, yielding exercise-level gap scores.
    The final recommendation set $\mathcal{E}_I$ is obtained by selecting the Top-$K$ exercises with the highest gap scores:
    \begin{equation}
    \mathcal{E}_I = \text{Top-}K \left( \mathbf{Q} \cdot \left( \mathbf{1} - \mathcal{M}_{\text{KT}}(\mathcal{H}_{s,t}) \right) \right).
    \end{equation}
    However, this formulation exhibits a locally greedy selection strategy, repeatedly recommending exercises associated with the most unmastered KCs, which leads to homogeneous recommendation results.
    To mitigate this issue, recent works adopt two-stage architectures that decouple candidate generation from ranking to promote recommendation diversity.
    Building upon KT-based retrieval, these paradigms incorporate a diversified re-ranking stage to cover a broader spectrum of KCs~\cite{WuLT2020, RenLS2023, LiuRG2025}.
    In the \textit{retrieval stage}, the algorithm identifies a candidate pool $\mathcal{E}_{\text{cand}}$ by selecting the $\text{Top-}N$ exercises that correspond to the student's estimated knowledge gaps.
    Subsequently, in the \textit{re-ranking stage}, the system selects an optimal subset $\mathcal{E}_I \subset \mathcal{E}_{\text{cand}}$ of size $K$ ($K < N$) that maximizes the set-level KC diversity:
    \begin{equation}
    \mathcal{E}_I = \mathop{\arg\max}_{\mathcal{E}' \subset \mathcal{E}_{\text{cand}}, |\mathcal{E}'|=K} \Omega_{div}(\mathcal{E}', \mathbf{Q}).
    \end{equation}
    The objective function $\Omega_{div}(\cdot, \mathbf{Q})$ promotes conceptual diversity by maximizing pairwise dissimilarity in KC coverage among the selected exercises.
    Both paradigms can be viewed under a unified framework of ILER.
    The output of an ILER method is a $\text{Top-}K$ subset of exercises:
    \begin{equation}
    \mathcal{E}_{I} = \{e_{1}, e_{2}, \dots, e_{K}\} \subset \mathcal{E}.
    \end{equation}
    Notably, the elements in $\mathcal{E}_{I}$ are treated as an \textit{unordered set}, without any explicit sequential relationship among the exercises.
    This drastically contrasts with path-level exercise recommendations, where the internal ordering and transition of exercises are explicitly modeled.
    
    \looseness=-1
    \noindent\textbf{Path-Level Exercise Recommendation (PLER).}
    The primary objective of PLER is to generate coherent learning paths that guide students toward specific pedagogical goals.
    As a representative work, SRC~\cite{CheSX2023} casts this task within a set-to-sequence paradigm, employing an encoder-decoder architecture to facilitate efficient and concept-aware path generation. 
    Formally, given the student's historical interaction sequence $\mathcal{H}_{s,t}$ and a target goal $G$, SRC aims to identify an optimal learning path $\mathcal{T}^*$ of length $N$.
    This task is framed within a full-path generation framework, where the entire sequence is generated to optimize global coherence:
    \begin{equation}
    \mathcal{T}^* = \text{Decoder} \left( \text{Encoder}(\mathcal{H}_{s,t}, G) \right) = \langle e_{p_1}, e_{p_2}, \dots, e_{p_N} \rangle,
    \end{equation}
    where $\text{Encoder}(\cdot)$ compresses both the target goals and an unordered set of unmastered knowledge concepts extracted from the historical sequence $\mathcal{H}_{s,t}$ into a latent global representation.
    Subsequently, $\text{Decoder}(\cdot)$ autoregressively maps this context into a fixed-length learning path within the exercise space $\mathcal{E}^N$.
    By adopting this full-path generation approach, the system is capable of modeling the underlying inter-exercise sequential dependencies and prerequisite constraints.
    
    \looseness=-1
    Alternatively, another prevalent approach within PLER formulates the task as a Markov Decision Process (MDP), predominantly leveraging Reinforcement Learning (RL) and Graph Neural Networks (GNN) to dynamically generate optimal learning paths~\cite{LiuTL2019, LiXY2023, WanCW2024, ZhaYP2024, CheLL2025}.
    Within these RL-based methods~\cite{KonT1999, TanCL2019, KubFM2021}, several advanced approaches explicitly incorporate knowledge structure graphs to model prerequisites among KCs, ensuring that foundational KCs are mastered before advancing to the target goal \cite{LiuTL2019, ZhaSX2024}.
    Formally, at each sequential step $i \in \{1, \dots, N\}$, the agent observes the current state, which comprises the target goal $G$ and the real-time knowledge state estimated from the dynamic interaction history $\mathcal{H}_{s, t+i-1}$, and selects the optimal exercise $e_{p_i}^*$ following a parameterized policy $\pi_{\Theta}$:
    \begin{equation}
    e_{p_i}^* = \mathop{\arg\max}_{e \in \mathcal{E}} \pi_{\Theta}(e \mid \mathcal{H}_{s, t+i-1}, G).
    \end{equation}
    Crucially, unlike static generation, the historical context expands dynamically after each intra-path recommendation, such that $\mathcal{H}_{s, t+i} = \mathcal{H}_{s, t+i-1} \cup \{(e_{p_i}^*, r_{p_i})\}$, where $r_{p_i}$ denotes the simulated instantaneous student feedback.
    Through these sequential state transitions and localized policy decisions, the algorithm synthesizes a strictly \emph{ordered sequence} of length $N$, designed to guide learners step-by-step:
    \begin{equation}
    \mathcal{T}^* = \langle e_{p_1}^*, e_{p_2}^*, \dots, e_{p_N}^* \rangle,
    \end{equation}
    where the internal transition from $e_{p_i}^*$ to $e_{p_{i+1}}^*$ is intrinsically driven by the maximization of long-term pedagogical outcomes.
    While these dynamic PLER methodologies demonstrate profound capabilities in formulating logically coherent paths, they have largely evolved as a siloed research domain. This divergence has resulted in a critical lack of cross-paradigm evaluation methods that can holistically benchmark both ILER and PLER within a unified recommendation framework.

    \begin{figure}[t]
        \centering
        \includegraphics[width=1\textwidth]{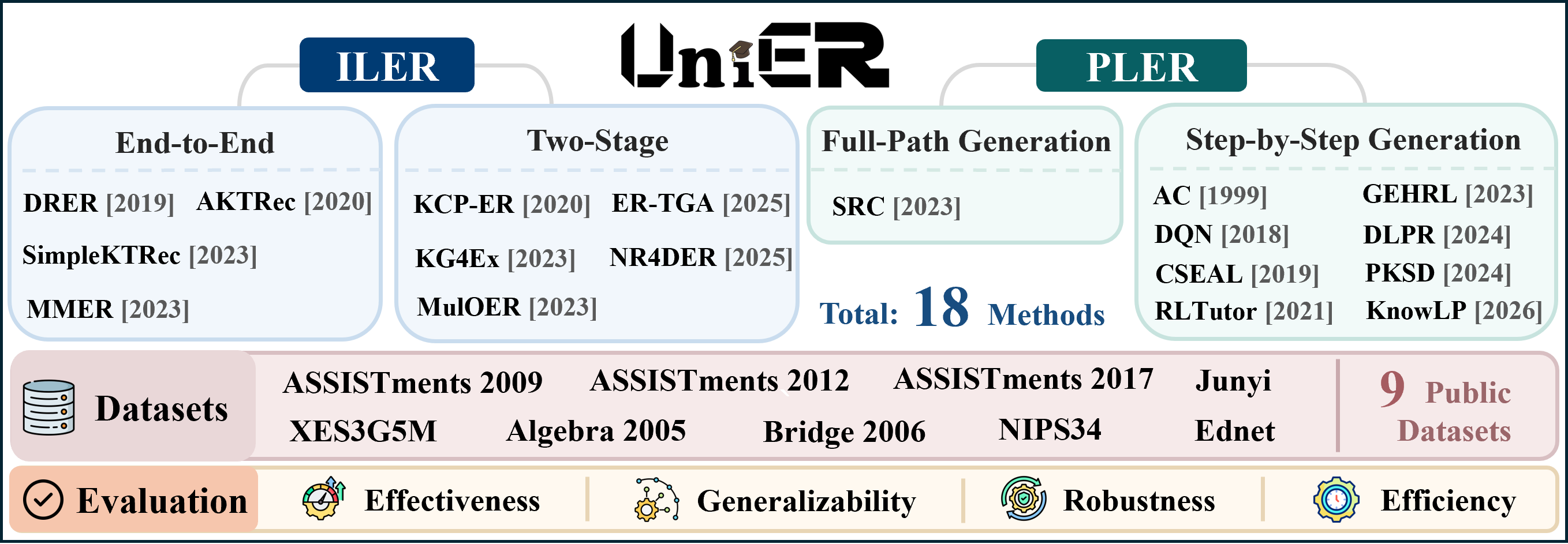}
        \caption{An overview of UniER.}
        \label{fig:overall framework}
    \end{figure}

    \section{Benchmark Design}
    \looseness=-1
    In this section, we provide a comprehensive overview of UniER covering the Weighted Cognitive Gain metric, algorithms, evaluation datasets, and implementation details.

    \subsection{Weighted Cognitive Gain Metric}
    \looseness=-1
    Evaluating the quality of recommended exercises is fundamental to personalized learning systems.
    However, existing evaluation metrics are highly fragmented and often focus on isolated aspects of performance, making it difficult to fairly compare different methodologies.
    More importantly, they implicitly assume that all knowledge concepts (KCs) are equally important, which contradicts real-world instructional settings where learning objectives vary across scenarios.
    In practice, certain KCs may play a more critical role depending on pedagogical goals, such as intensive exam-oriented training versus broad knowledge exploration.
    To address this limitation, we propose Weighted Cognitive Gain (WCG), a configurable evaluation metric that allows flexible weighting over KCs. By explicitly modeling concept importance, WCG enables scenario-aware evaluation and provides a unified framework for fair comparison across different recommendation paradigms.
    
    \looseness=-1
    \noindent\textbf{Unified Output.}
    To enable a fair and strictly aligned quantitative evaluation, it is essential to establish a unified output. 
    Existing paradigms produce structurally divergent outputs: PLER directly generates an ordered learning path $\mathcal{T}^*$ of length $N$, whereas ILER yields an unordered set $\mathcal{E}_I$.
    To bridge this structural dichotomy, we project the ILER output into a sequential learning path by sorting the exercises in $\mathcal{E}_I$ in descending order of their predicted recommendation scores.
    Consequently, we establish a unified output space: regardless of the underlying paradigm, the evaluated method must ultimately formulate an ordered learning path $\mathcal{T}_{\text{unified}}$ constrained by a cognitive budget of $N$ steps.
    
    \looseness=-1
    \noindent\textbf{Unified Evaluation Metric.}
    To establish a rigorously fair benchmark, we formalize a unified metric: WCG.
    We introduce a task-specific target weight distribution $\mathbf{w} \in [0, 1]^{|\mathcal{C}|}$, subject to $\sum_{c \in \mathcal{C}} w_c = 1$.
    Under this unified evaluation framework, the performance of any generated learning path $\mathcal{T}_{\text{unified}}$ is quantified by measuring its expected weighted knowledge mastery improvement:
    \begin{equation} 
    \text{WCG}(\mathcal{T}_{\text{unified}} \mid \mathbf{w}) = \sum_{c \in \mathcal{C}} w_c \cdot \left( \mathcal{M}^{(c)}_{\text{KT}}(\mathcal{H}_{s,t} \oplus \text{Sim}(\mathcal{T}_{\text{unified}}))-\mathcal{M}^{(c)}_{\text{KT}}(\mathcal{H}_{s,t}) \right),
    \end{equation}
    where $\mathcal{M}^{(c)}_{\text{KT}}(\mathcal{H}_{s,t})$ and $\mathcal{M}^{(c)}_{\text{KT}}(\mathcal{H}_{s,t}\oplus\text{Sim}(\mathcal{T}_{\text{unified}}))$ denote the student's estimated mastery level on concept $c$ before and after executing the learning path, respectively.
    Here, $\oplus$ denotes sequence concatenation, and $\text{Sim}(\cdot)$ represents the simulated response generation process for the recommended path, executed by the underlying knowledge state simulator.
    The weight configuration $\mathbf{w}$ can be flexibly adjusted to accommodate specific pedagogical requirements, a key advantage that enables UniER to generalize across diverse educational contexts.
    In this study, we primarily instantiate two representative settings to demonstrate this adaptability in real-world scenarios:

    \begin{itemize}[leftmargin=*]
        \item \textbf{Task 1: Targeted Goal Achievement (TGA).} Let $\mathcal{C}_{\text{target}} \subset \mathcal{C}$ denote a specific set of target concepts a student needs to master. We construct a highly sparse weight vector by defining $w_c = 1 / |\mathcal{C}_{\text{target}}|$ if $c \in \mathcal{C}_{\text{target}}$, and $w_c = 0$ otherwise. This setting strictly evaluates the model's architectural capacity to synthesize logically progressive paths that fulfill specific pedagogical goals, effectively quantifying the average cognitive gain across the designated targets. In practical, TGA aligns with the pedagogical scenario of targeted exam preparation, where a student must efficiently master a predefined set of crucial KCs.
        
        \item \textbf{Task 2: Global Proficiency Promotion (GPP).} Let $\mathcal{C}_{\text{unmastered}} \subset \mathcal{C}$ denote the set of all concepts the student has yet to master at time $t$. We define a uniform distribution over these concepts by setting $w_c = 1 / |\mathcal{C}_{\text{unmastered}}|$ if $c \in \mathcal{C}_{\text{unmastered}}$, and $w_c = 0$ otherwise. This task evaluates the model's capability to maximize overall knowledge coverage and broad deficit remediation. Practically, this corresponds to the pedagogical scenario of comprehensive end-of-term reviews, where the primary objective is to elevate the student's overall academic proficiency.
    \end{itemize}

    \subsection{Benchmark Algorithms}
    To ensure a comprehensive and representative evaluation, UniER incorporates a diverse suite of state-of-the-art ER methods.
    These algorithms are categorized into two primary paradigms based on their recommendation granularity and structural design: Item-Level Recommendation (ILER) and Path-Level Recommendation (PLER).
    For a concise summary of the specific model configurations and implementation details, please refer to Figure \ref{fig:overall framework} and Appendix \ref{Detailed Description of Algorithms in UniER}.
    
    \subsection{Datasets and Implementation Details}
    \looseness=-1
    \noindent\textbf{Datasets.}
    We evaluate UniER on nine widely recognized datasets: ASSISTments (2009, 2012, and 2017)~\cite{AbdW2019}, Bridge2006, Junyi~\cite{LiXY2024}, XES3G5M~\cite{LiuLG2023}, Algebra2005~\cite{StaNS2010}, Ednet, and NIPS34~\cite{LiuLC2022}, which span a diverse array of educational domains and scales.
    Comprehensive data statistics and preprocessing specifications are detailed in Appendix \ref{Detailed Description of Datasets in UniER}.

    \looseness=-1
    \noindent\textbf{Hyperparameter Search.}
    To obtain the performance upper bounds of various methods on ILER/PLER tasks, we conduct a random search to find the optimal hyperparameters.
    The search space is detailed in Tables \ref{tab:hyper-parameter-ILER} and \ref{tab:hyper-parameter-PLER}.
    The random search is conducted 20 times or for a maximum of one day per method per dataset to ensure fairness.
    For more details related to the experimental setup in UniER, please refer to Appendix \ref{Additional Experimental Details}.
       
    \section{Experimental Results \& Discussion}
    In this section, we present a comprehensive evaluation and in-depth analysis on the UniER benchmark. Specifically, our experiments are designed to answer the following core research questions (RQs):
    \begin{itemize}[leftmargin=*, itemsep=0pt, parsep=0pt, topsep=0pt]
    \item \textbf{RQ1 (Effectiveness):} How do ILER and PLER perform under the unified WCG metric?
    \item \textbf{RQ2 (Generalizability):} How does performance degrade under data sparsity and cold-start?
    \item \textbf{RQ3 (Robustness):} How robust are these representative models against underlying label noise?
    \item \textbf{RQ4 (Efficiency):} What is the trade-off between computational cost and pedagogical performance?
    \end{itemize}

\begin{table}[t]
\centering
\caption{Comparison in terms of GPP@10 and TGA@10. The best three results are highlighted by \colorbox{gold}{1st}, \colorbox{silver}{2nd}, and \colorbox{bronze}{3rd}. ``Avg.'' indicate the average GPP@10 and TGA@10 across all datasets.}
\label{tab:overall performance_WCG}
\scriptsize
\setlength{\tabcolsep}{3pt}
\begin{tabular}{l|cc|cc|cc|cc|cc|cc|cc|c}
\toprule
\multirow{2}{*}{} & \multicolumn{2}{c|}{ASSIST2017} & \multicolumn{2}{c|}{Algebra2005} & \multicolumn{2}{c|}{Bridge2006} & \multicolumn{2}{c|}{Ednet} & \multicolumn{2}{c|}{Junyi} & \multicolumn{2}{c|}{NIPS34} & \multicolumn{2}{c|}{XES3G5M} & \multicolumn{1}{c}{Avg.}\\
\cmidrule(lr){2-3} \cmidrule(lr){4-5} \cmidrule(lr){6-7} \cmidrule(lr){8-9} \cmidrule(lr){10-11} \cmidrule(lr){12-13} \cmidrule(lr){14-15} \cmidrule(lr){16-16}
\rowcolor{gray!20}
 & TGA & GPP & TGA & GPP & TGA & GPP & TGA & GPP & TGA & GPP & TGA & GPP & TGA & GPP & T\&G  \\
\midrule
\rowcolor{red!10}
DRER  &-0.104 &-0.009 &-0.027 &0.024 &-0.005 &-0.001 &-0.162 &-0.065 &0.014 &-0.098 &0.050 &0.029 &0.124 &0.003 &-0.019  \\
\rowcolor{red!10}
AKTRec  &-0.123 &-0.072 &0.049 &0.023 &0.000 &0.000 &-0.483 &-0.155 &-0.067 &-0.145 &0.018 &-0.032 &0.134 &-0.017 &-0.049  \\
\rowcolor{red!10}
S-KTRec  &-0.123 &-0.067 &0.023 &0.036 &0.000 &0.000 &-0.454 &-0.142 &-0.082 &-0.133 &0.024 &0.035 &0.125 &-0.018 &-0.057  \\
\rowcolor{red!10}
MMER  &-0.128 &-0.121 &0.063 &0.049 &-0.008 &-0.004 &-0.229 &-0.121 &0.062 &-0.005 &0.032 &0.045 &0.190 &0.019  &-0.015 \\
\midrule
\rowcolor{blue!10}
KCP-ER  &-0.160 &-0.104 &0.046 &0.032 &0.000 &0.001 &-0.012 &0.002 &0.064 &0.054 &0.146 &0.041 &0.168 &0.020 &0.033  \\
\rowcolor{blue!10}
KG4Ex  &-0.208 &-0.116 &0.116 &0.046 &-0.001 &-0.002 &-0.119 &-0.128 &0.011 &-0.061 &-0.087 &-0.087 &0.085 &-0.075 &-0.054 \\
\rowcolor{blue!10}
MulOER  &-0.114 &-0.088 &0.065 &0.034 &-0.005 &-0.002 &-0.086 &-0.059 &0.027 &0.009 &0.045 &0.008 &0.195 &0.054 &0.016 \\
\rowcolor{blue!10}
ER-TGA  &-0.117 &-0.097 &0.107 &0.051 &-0.003 &0.004 &-0.101 &-0.125 &0.068 &0.118 &-0.190 &-0.124 &0.063 &-0.021 &-0.035 \\
\rowcolor{blue!10}
NR4DER  &-0.139 &-0.089 &0.088 &0.042 &0.002 &0.003 &-0.352 &-0.153 &0.019 &-0.179 &0.084 &-0.002 &0.162 &-0.018 &-0.032 \\
\midrule
\rowcolor{green!10}
SRC &0.024 &-0.135 &0.167 &0.082 &0.000 &\cellcolor{gold}{0.007} &\cellcolor{silver}{0.306} &-0.340 &0.102 &-0.103 &0.116 &0.170 &-0.026 &0.017 &0.011  \\
\midrule
\rowcolor{yellow!10}
AC & \cellcolor{gold}{0.283} &0.179 & \cellcolor{gold}{0.244} &0.096 &0.000 &0.005 &-0.38 &0.405 &\cellcolor{gold}{0.457} &0.355 &0.419 &\cellcolor{silver}{0.485} &0.902 &-0.246 &0.190 \\
\rowcolor{yellow!10}
DQN &0.164 &0.121 &0.084 &\cellcolor{gold}{0.126} &0.004 &\cellcolor{silver}{0.006} &0.092 &\cellcolor{gold}{0.486} &0.212 &0.284 &\cellcolor{bronze}{0.509} &0.105 &0.877 &\cellcolor{silver}{0.255} &0.220 \\
\rowcolor{yellow!10}
CSEAL &0.132 &\cellcolor{bronze}{0.208} &0.162 &0.054 &-0.001 &0.005 &0.042 &0.363 &\cellcolor{silver}{0.419} &\cellcolor{bronze}{0.370} &0.400 &\cellcolor{bronze}{0.481} &\cellcolor{silver}{0.914} &-0.118 &0.210 \\
\rowcolor{yellow!10}
RLTutor &\cellcolor{bronze}{0.226} &\cellcolor{silver}{0.246} &\cellcolor{bronze}{0.213} &\cellcolor{silver}{0.125} &\cellcolor{bronze}{0.005} &0.005 &-0.181 &\cellcolor{bronze}{0.469} &\cellcolor{bronze}{0.318} &\cellcolor{silver}{0.378} &\cellcolor{silver}{0.617} &0.443 &0.888 &-0.126 &\cellcolor{silver}{0.259}  \\
\rowcolor{yellow!10}
GEHRL &-0.226 &-0.144 &0.161 &0.034 &\cellcolor{gold}{0.017} &\cellcolor{gold}{0.007} &\cellcolor{gold}{0.329} &\cellcolor{silver}{0.484} &-0.277 &-0.214 &-0.17 &0.101 &\cellcolor{gold}{0.918} &\cellcolor{bronze}{0.243}  &0.182 \\
\rowcolor{yellow!10}
DLPR &0.093 &0.140 &0.181 &0.112 &\cellcolor{silver}{0.006} &\cellcolor{silver}{0.006} &-0.109 &0.364 &0.218 &0.275 &0.322 &0.357 &0.755 &0.237  &0.179 \\
\rowcolor{yellow!10}
PKSD &\cellcolor{silver}{0.268} &\cellcolor{gold}{0.290} &\cellcolor{silver}{0.235} &\cellcolor{silver}{0.125} &0.001 &\cellcolor{silver}{0.006} &\cellcolor{bronze}{0.225} &0.417 &0.288 &\cellcolor{gold}{0.381} &\cellcolor{gold}{0.644} &\cellcolor{gold}{0.517} &\cellcolor{silver}{0.914} &0.152  &\cellcolor{gold}{0.311}\\
\rowcolor{yellow!10}
KnowLP &-0.031 &0.171 &0.203 &0.123 &0.004 &0.005 &0.107 &0.370 &0.255 &0.355 &0.474 &0.381 &0.803 &\cellcolor{gold}{0.256}  &\cellcolor{bronze}{0.232} \\
\bottomrule
\end{tabular}
\end{table}

    \subsection{Overall Performance Comparison (RQ1)}
    \looseness=-1
    \noindent\textbf{Experiment Design.} We comprehensively compare the exercise recommendation performance of 18 baseline algorithms across 9 benchmark datasets in terms of GPP@10 and TGA@10, where @10 denotes that the evaluation is conducted over a learning path of 10 recommendation steps.
    
    \looseness=-1
    \noindent\textbf{Experimental Results.} Table \ref{tab:overall performance_WCG} presents the overall performance comparison in terms of TGA and GPP.
    Based on these comprehensive results, we summarize the following key observations:
    
    \looseness=-1
    \noindent\textbf{\underline{Observation \ding{182}}:} \textbf{Overall superiority of PLER methods.}
    Table \ref{tab:overall performance_WCG} shows a clear performance gap between the two approaches. In the ``Avg. T\&G'' column, most ILER methods result in negative average scores. Even the best-performing ILER models, such as MulOER (0.016) and KCP-ER (0.033), only achieve marginal positive gains. In contrast, RL-based PLER models situated at the bottom block maintain substantial positive gains, with PKSD reaching the highest average of 0.311. Furthermore, even SRC achieves a positive average gain (0.011) that remains highly competitive with the most complex ILER models. This fundamentally proves that adopting a long-term path planning perspective, rather than just predicting the next step, is the key to improving overall learning outcomes.
    
    \looseness=-1
    \noindent\textbf{\underline{Observation \ding{183}}:} \textbf{Context-specific TGA-GPP trade-offs.}
    The results show that improving TGA does not always lead to improvements in GPP. For instance, on the XES3G5M dataset, the AC model achieves a high TGA (0.902) but a negative GPP (-0.246). This trade-off implies that focusing too heavily on specific targets can sometimes harm a student's broader knowledge base. In practice, this means we should choose models based on the learning scenario: models with high TGA are particularly effective for short-term targeted learning, while balanced models are better for building long-term, comprehensive skills.
    
    \looseness=-1
    \noindent\textbf{\underline{Observation \ding{184}}:} \textbf{No universally superior method.}
    Table \ref{tab:overall performance_WCG} demonstrates that no single method consistently outperforms others on all datasets. While PKSD has the best average score, it is beaten on specific datasets. For example, on the Ednet dataset, GEHRL (TGA 0.329, GPP 0.484) easily outperforms PKSD (TGA 0.225, GPP 0.417). Similarly, AC obtains a higher TGA on ASSIST2017 (0.283) than PKSD (0.268). This shows that how well a model performs depends heavily on the specific characteristics of the dataset, meaning we need to select algorithms adaptively for different real-world platforms.
    
    \looseness=-1
    \noindent\textbf{\underline{Observation \ding{185}}:} \textbf{Performance collapse from missing pedagogical logic.}
    ILER models show severe weaknesses on large datasets with long interaction histories, such as Ednet and Junyi. On Ednet, for instance, AKTRec scores -0.483 for TGA and -0.155 for GPP. Because ILER models only focus on finding the immediate next best exercises. They entirely ignore how the student's knowledge evolves while working through this list. As a result, the recommended exercises often lack a logical learning order or contain redundant concepts, leading to poor overall learning outcomes. PLER models avoid this issue by explicitly planning a logical path, ensuring that each exercise thoroughly prepares the student for the next one.
\begin{figure*}[t]
    \centering
    \includegraphics[width=1\textwidth]{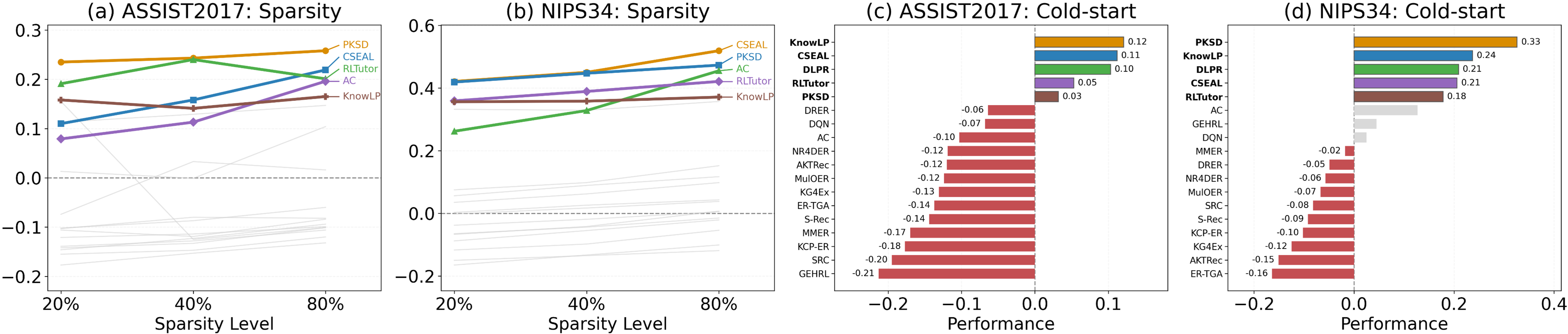}
    \caption{
     Generalizability analysis under data sparsity and cold-start settings. In subfigures (a) and (b), the top-five methods are highlighted, while the remaining methods are represented by gray lines.
    }
    \label{fig:RQ2}
\end{figure*}

    \subsection{Performance under Data Sparsity and Cold-Start (RQ2)}
    \looseness=-1
    \noindent\textbf{Experiment Design.} To assess the generalizability of ILER/PLER methods, we simulate two extreme yet common educational scenarios on the ASSIST2017 and NIPS34 datasets: \textit{data sparsity} and \textit{cold-start}. For data sparsity, we evaluate performance under three varying sparsity levels (20\%, 40\%, and 80\%). For the cold-start scenario, we test the models on a specialized subset where each student's historical interaction sequence is truncated to only five steps.

    \looseness=-1
    \noindent\textbf{Experimental Results.} Figure \ref{fig:RQ2} illustrates the performance comparison (evaluated via GPP@10) under varying sparsity levels (panels a and b) and cold-start conditions (panels c and d). Beyond the general failure of ILER models, analyzing the performance of PLER models reveals two critical insights regarding model generalizability:

    \looseness=-1
    \noindent\textbf{\underline{Observation \ding{186}}:} \textbf{Performance consistency under extreme sparsity.}
    The sparsity experiments reveal a stark contrast in generalizability. As dataset density decreases from 80\% to 20\% on NIPS34, ILER models experience severe performance drops deep into negative territory (\eg, the GPP@10 of ER-TGA falls from -0.119 to -0.150). In contrast, top PLER models maintain high consistency. For instance, PKSD only experiences a minor decrease in GPP@10 from 0.473 to 0.419, retaining nearly 90\% of its performance despite losing most of the interaction logs. This indicates that planning along sequential learning paths, rather than relying on isolated exercise predictions, allows PLER to maintain strong generalizability even when data is severely incomplete.
    
    \looseness=-1
    \noindent\textbf{\underline{Observation \ding{187}}:} \textbf{SOTA reshuffling under individual cold-start.}
    The results expose a fundamental difference between data sparsity and cold-start. Generalizing well to dataset sparsity does not guarantee few-shot transferability. For example, on ASSIST2017, the AC model handles sparsity well but collapses to -0.103 under the 5-step cold-start setting. Furthermore, this extreme setting completely reshuffles the top performers: PKSD, the overall SOTA, drops drastically to a marginal GPP@10 of 0.032 on ASSIST2017, while KnowLP takes the lead (0.121). This combined phenomenon demonstrates that integrating student-independent structural priors (\eg, knowledge structure graphs) is crucial for ensuring generalizability.

\begin{figure}[t]
    \centering
    \includegraphics[width=1\columnwidth]{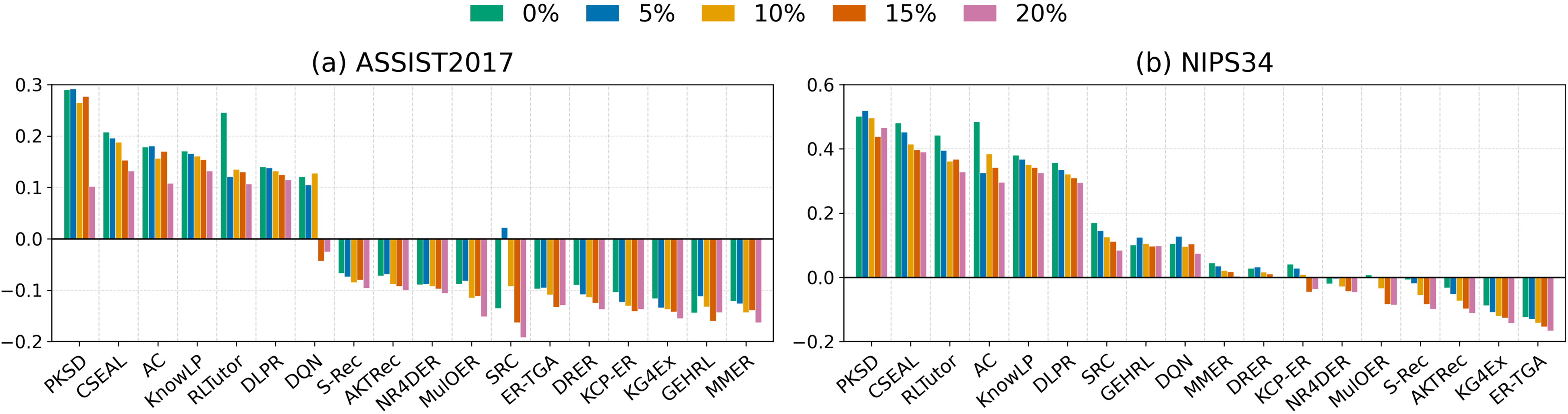}
    \caption{Performance under different perturbation levels.}
    \label{fig:robustness}
\end{figure}

    \subsection{Stability Against Label Noise (RQ3)}
    \looseness=-1
    \noindent\textbf{Experiment Design.} To evaluate model performance in the presence of real-world data noise, we randomly alter the correctness labels in student records at ratios of 5\%, 10\%, 15\%, and 20\%.

   \looseness=-1
    \noindent\textbf{Experimental Results.} Figure \ref{fig:robustness} shows the performance changes (evaluated via GPP@10) compared to the clean dataset (0\% noise). The results reveal one key finding about model robustness:

   \looseness=-1
    \noindent\textbf{\underline{Observation \ding{188}}:} \textbf{Structurally buffered noise tolerance.} The results expose a severe structural vulnerability in how models process local perturbations. As label noise reaches 20\%, ILER methods suffer rapid performance degradation (\eg, DRER's GPP@10 on ASSIST2017 falls from -0.090 to -0.137). Because ILER relies on isolated predictions from a single frozen state, a noisy historical label directly corrupts the entire static batch. Conversely, top PLER models remain highly stable (\eg, PKSD on NIPS34 only sees a minor decrease from 0.502 to 0.466). By constructing learning paths grounded in inherent pedagogical logic, PLER possesses a structural error-buffering capacity, utilizing the coherent dependencies among exercises to absorb isolated noisy signals rather than catastrophically failing from them.

\begin{figure}[t]
\centering
\includegraphics[width=1\columnwidth]{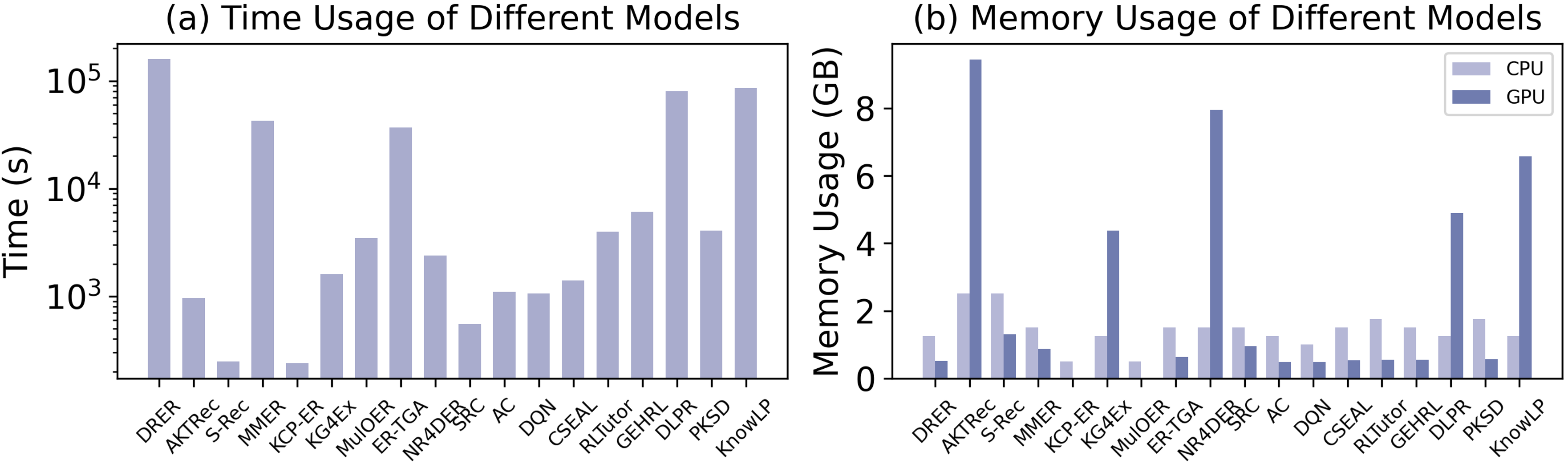}
\caption{Time and memory usage comparison.}
\label{fig:time_memory_usage}
\end{figure}

\subsection{Efficiency Analysis (RQ4)}
   \looseness=-1
    \noindent\textbf{Experiment Design.} We evaluate the computational efficiency of ILER/PLER methods using default hyperparameter settings on ASSIST2017. Our assessment focuses on two main aspects: time efficiency and memory usage.

   \looseness=-1
    \noindent\textbf{Experimental Results.} The computational overhead of each method, quantified by total time consumption and memory usage, is detailed in Figures \ref{fig:time_memory_usage} (a) and (b), respectively. Based on these findings, we highlight a striking observation:
    
   \looseness=-1
    \noindent\textbf{\underline{Observation \ding{189}}:} \textbf{Extreme intra-paradigm scalability variance.}
    Figure \ref{fig:time_memory_usage}, computational overhead varies drastically even among RL-based models. Heavy-architecture RL-based methods (\eg, DLPR and KnowLP) suffer from severe scalability bottlenecks, requiring tens of hours ($>10^5$s) to converge and consuming massive GPU memory due to the exorbitant cost of searching and reasoning over integrated knowledge structure graphs. Conversely, streamlined RL-based methods (\eg, AC, DQN, and CSEAL) efficiently reach convergence in a fraction of the time ($\sim 10^3$s) while maintaining a lightweight memory footprint ($<1$ GB). This reveals that the true computational bottleneck stems from coupling expensive structural graph searches with policy optimization, rather than the intrinsic sampling cost of the RL paradigm itself.

    \section{Conclusion and Future Directions}
    In this paper, we present a comprehensive benchmark to evaluate existing methods under various settings, including sparsity and cold-start scenarios.
    While current approaches demonstrate promising performance, our study also reveals several fundamental limitations in both problem formulation and evaluation protocols, which call for further investigation.
    \textbf{First}, the inherent unobservability of latent knowledge states remains a critical challenge. Existing evaluation protocols largely rely on next-item performance as proxy labels. However, this correlational mapping fails to accurately isolate the true pedagogical gain caused by a specific recommendation path, making it difficult to assess true causal effectiveness.
    \textbf{Second}, while our benchmark unifies the evaluation on existing public data, there is a fundamental mismatch in data collection paradigms. Most datasets are passively collected for Knowledge Tracing (KT), lacking the exploratory, multi-step sequential decisions required to fully validate path-level recommendation quality.
    \textbf{Finally}, there is no unified formulation of real-world educational objectives. Different studies adopt inconsistent assumptions regarding learning goals, such as mastery improvement, efficiency, or diversity, leading to fragmented evaluation criteria and incomparable results across works.
    Based on these observations, we highlight several important directions for future research:
    
    \looseness=-1
    \textbf{(1) Toward causal and interactive evaluation signals.}
    Since absolute ground truth for knowledge states is inherently inaccessible in offline logs, future work should pivot towards causal inference techniques (\eg, counterfactual estimation) or leverage well-calibrated simulated student environments to better approximate true learning gains.
    
    \looseness=-1
    \textbf{(2) Paradigm shift in data collection.}
    Moving forward, the community must construct dedicated datasets collected through active, sequential educational interventions, rather than relying solely on observational KT logs.
    
    \looseness=-1
    \textbf{(3) Unified modeling of educational objectives.}
    Future research should aim to formalize and unify different educational goals into a coherent framework, enabling fair comparison and more principled method design.
    
    \looseness=-1
    Overall, our findings highlight that advancing exercise recommendation requires not only better models, but also a deeper rethinking of problem formulation, data construction, and evaluation protocols.
    
    \looseness=-1
    \noindent\textbf{Limitation.} While providing a unified evaluation framework, our benchmark has two primary limitations.
    First, our current protocol is grounded in \textbf{offline evaluation}, which lacks the dynamic, closed-loop feedback inherent in actual tutoring scenarios. Consequently, it cannot fully capture how models adapt to real-time student behavioral shifts. 
    Second, regarding \textbf{metrics}, the observed trade-offs (\eg, TGA vs. GPP) indicate that our framework currently lacks a singular, universally aligned metric to resolve conflicting pedagogical objectives. 
    As an evolving benchmark, future versions will address these gaps by incorporating interactive simulation protocols and developing multi-objective optimization metrics, supported by the integration of more diverse, task-specific datasets.

{
\small
\bibliographystyle{IEEEtran}
\bibliography{UniER}
}

\newpage
\appendix
\section{Detailed Description of Datasets in UniER} \label{Detailed Description of Datasets in UniER}
The description of datasets in UniER in given as follows:
    \begin{itemize}[leftmargin=*]
        \item \textbf{ASSISTments 2009 / 2012 / 2017.} These are benchmark knowledge‑tracing datasets collected from the ASSISTments online tutoring platform. The 2009 and 2012 datasets represent different school years, whereas the 2017 subset originates from the ASSISTments Data Mining Competition.
        \item \textbf{Algebra2005.} From the KDD Cup 2010 educational data mining challenge, this dataset contains detailed algebra exercise logs with knowledge concept annotations. It captures responses from middle school students practicing multi‑step algebra problems.
        \item \textbf{Bridge2006.} Also a cornerstone of the KDD Cup 2010 challenge, the Bridge2006 dataset captures an extensive array of student interactions with algebra exercises. Its massive interaction volume offers the statistical robustness required to benchmark models on long-term performance prediction and complex temporal dynamics.
        \item \textbf{NIPS34.} Originating from the NeurIPS 2020 Education Challenge, this dataset comprises diagnostic math questions annotated with leaf-node skill labels for each item. Its unique structure is particularly effective for evaluating the robustness and generalization of knowledge tracing models when confronted with high question sparsity.
        \item \textbf{Junyi.} Collected from Junyi Academy, the Junyi dataset contains student problem-solving logs, exercise-level metadata, and annotated relationships among exercises. 
        \item \textbf{Ednet.} A large‑scale hierarchical educational dataset collected by the Santa AI tutoring platform. It records a rich variety of student activities, ranging from exercise solving to content browsing, over more than two years, making it one of the largest interactive educational system datasets publicly available.
        \item \textbf{XES3G5M.} This benchmark, recently released at NeurIPS 2023, is derived from a real-world K-12 online math learning environment. Beyond standardized correctness sequences, it encompasses an enriched set of auxiliary metadata, including hierarchical knowledge concepts and the textual attributes of questions."
    \end{itemize}

    \begin{table}[ht]
    \centering
    \caption{Statistics of the datasets used in our benchmark.}
    \label{tab:dataset_stats}
    \begin{tabular}{lrrrr}
    \toprule
    \textbf{Dataset} & \textbf{\#Interactions} & \textbf{\#Students} & \textbf{\#Exercises} & \textbf{\#KC} \\
    \midrule
    ASSISTments2009 & 346,860 & 4,217 & 26,688 & 123 \\
    ASSISTments2012 & 6,123,270 & 46,674 & 179,999 & 265 \\
    ASSISTments2017 & 942,816 & 686 & 3,162 & 102 \\
    Algebra2005 & 813,661 & 575 & 1,084 & 112 \\
    Bridge2006 &3,686,871 & 1,146 & 19,258 & 493 \\
    NIPS34 & 1,382,727 & 4,918 & 948 & 57 \\
    Junyi  & 25,925,992 & 247,606 & 722 & 41 \\
    Ednet &95,293,926 & 784,309 & 13,169 & 188 \\
    XES3G5M & 5,549,635 & 18,066 & 7,652 & 865 \\
    \bottomrule
    \end{tabular}
    \end{table}

\section{Detailed Description of Algorithms in UniER} \label{Detailed Description of Algorithms in UniER}
The description of benchmarking algorithms in UniER is demonstrated as follows.

\noindent\textbf{Type I: End-to-end.}
    \begin{itemize}[leftmargin=*]
        \item \textbf{DRER.} The paper proposes DRER, a deep reinforcement learning framework for adaptive exercise recommendation. It introduces Exercise Q-Networks to capture students’ exercising states and histories, and designs multi-objective rewards to jointly optimize Review \& Explore, Smoothness, and Engagement.
        \item \textbf{AKTRec.} The AKT model introduces a context-aware attentive knowledge tracing framework that uses two self-attentive encoders and a monotonic attention mechanism to capture a student’s past exercise history. Building on AKT, AKTRec identifies knowledge concepts that a student has not yet mastered and recommends exercises corresponding to these gaps, enabling personalized learning.
        \item \textbf{SimpleKTRec.} SimpleKT introduces a lightweight attention-based knowledge tracing framework. It represents each exercise embedding as a combination of its associated knowledge concepts and leverages a simple dot-product attention mechanism to extract temporal knowledge states efficiently. Building on SimpleKT, SimpleKTRec identifies unmastered knowledge concepts for each student and recommends corresponding exercises, supporting personalized learning.
        \item \textbf{MMER.} The paper proposes MMER, a meta multi-agent reinforcement learning framework for exercise recommendation, where each knowledge concept is treated as an agent with competitive and cooperative interactions. A meta-training stage enables the model to quickly adapt to new student groups with few-shot data, while the multi-agent design allows efficient long-term learning of KCs.
    \end{itemize}
    
\noindent\textbf{Type II: Two-stage.}
    \begin{itemize}[leftmargin=*]
        \item \textbf{KCPER.} KCP-ER adopts a two-stage architecture for exercise recommendation, comprising a Knowledge Concept Prediction Layer (KCPL) and an Exercise Set Filtering Layer (ESFL). KCPL predicts the coverage and mastery of knowledge concepts using LSTM for coverage and DKT for mastery, generating vectors that represent both student knowledge state and the likelihood of each concept appearing in future exercises. Based on these predictions, ESFL first filters exercises to ensure suitable difficulty and novelty, and then uses a simulated annealing-based list generator to maximize diversity, producing a recommended exercise list.
        \item \textbf{KG4Ex.} KG4Ex employs a two-stage architecture consisting of a feature extraction module and a knowledge graph-based recommendation module. The feature extraction module uses LSTM and exponential functions to compute students’ mastery levels, the probability of knowledge concepts appearing in the next exercise, and forgetting rates. Based on these features, the knowledge graph module constructs relationships among students, exercises, and knowledge concepts, learns embeddings with TransE, TransE-adv, or RotatE~\cite{WanKL2024, WanWG2024}, and recommends suitable exercises while providing explainable recommendation reasons~\cite{GuaXC2023, GuaCX2025}.
        \item \textbf{MulOER.} MulOER-SAN adopts a two-layer multi-objective framework for exercise recommendation, where the bottom layer uses self-attention networks to predict knowledge concept coverage and dynamically trace students’ mastery levels. The top layer filters exercises from the candidate subset, applying a chaotic sparrow search algorithm to maximize diversity while a smoothness factor ensures gradual difficulty progression.
        \item \textbf{ER-TGA.} ER-TGA adopts a two-stage framework combining cognitive diagnosis and a tribal-alliance genetic algorithm for exercise filtering. The cognitive diagnosis layer identifies students’ weak knowledge components and constructs a candidate exercise set, addressing cold-start scenarios. Building on this, TGA selects the optimal subset of exercises minimizing a multi-objective recommendation cost (accuracy, novelty, diversity, proximity, coverage, quantity, and volatility), producing personalized exercise recommendations aligned with learners’ knowledge gaps.
        \item \textbf{NR4DER.} NR4DER employs a two-stage framework consisting of an exercise filter module and a neural re-ranking module. The exercise filter predicts students’ knowledge concept mastery using an mLSTM-based predictor and constructs candidate exercises with appropriate difficulty, enhanced for inactive students via a student representation enhancer. Building on this, the neural re-ranking module integrates exercise relevance and learning pattern diversity, generating personalized and diversified exercise recommendation lists that align with individual students’ learning pace.
    \end{itemize}

\noindent\textbf{Type III: Full-path.}
    \begin{itemize}[leftmargin=*]
        \item \textbf{SRC.} SRC formulates learning path recommendation as a set-to-sequence task, where a concept-aware encoder captures correlations among candidate learning concepts and produces global concept representations. An attention-based decoder sequentially generates the learning path, guided by predicted student mastery and inter-concept dependencies. A knowledge-tracing auxiliary module predicts mastery of each concept at every step, stabilizing training and improving the quality of generated learning paths.
    \end{itemize}

\noindent\textbf{Type IV: Step-by-step.}
    \begin{itemize}[leftmargin=*]
        \item \textbf{AC.} The classical Actor-Critic Algorithms framework is applied to step-by-step learning path recommendation, where the critic estimates the expected value of the current student state and the actor selects the next exercise to optimize learning outcomes. Temporal-difference learning stabilizes the critic’s estimation, while the actor updates policy parameters along the gradient of the expected reward, allowing dynamic adaptation to students’ evolving knowledge. This framework generates personalized learning paths sequentially, making decisions at each step based on current student knowledge and prior interactions.
        \item \textbf{DQN.} DQN applies a deep convolutional neural network to reinforcement learning, where the agent observes high-dimensional pixel input and sequentially selects actions to maximize future rewards. The network is trained using a variant of Q-learning with experience replay, which stabilizes learning by breaking correlations between sequential data and enables off-policy updates. At each step, the agent updates its estimate of the Q-values for possible actions based on observed rewards, effectively learning step-by-step control policies that adapt to evolving states and sequential dependencies.
        \item \textbf{CSEAL.} CSEAL models the learning path recommendation as a Markov Decision Process, combining knowledge tracing, cognitive navigation, and an actor-critic recommender to generate exercises sequentially. Knowledge Tracing estimates evolving student knowledge mastery levels, while the Cognitive Navigation module selects candidate items according to the prerequisite graph to ensure logical consistency and reduce search space.
        \item \textbf{RLTutor.} RLTutor models adaptive tutoring as a step-by-step sequential decision problem, where an Inner Model estimates the student’s current knowledge state using the DAS3H knowledge tracing method, incorporating temporal memory dynamics and prior learning history. A reinforcement learning agent (PPO-based) optimizes the teaching strategy for the Inner Model, generating the next exercise recommendation while minimizing interactions with actual students.
        \item \textbf{GEHRL.} GEHRL employs a hierarchical reinforcement learning framework for goal-oriented learning path recommendation, with a high-level agent selecting sub-goals and a low-level agent recommending learning items sequentially to achieve each sub-goal. A graph-based candidate selector constrains the low-level agent’s action space to ensure the path includes only goal-related exercises, improving learning efficiency.
        \item \textbf{DLPR.} DLPR addresses learning path recommendation as a step-by-step sequential decision problem, generating paths dynamically based on real-time student interactions and mastery feedback. It employs a Difficulty-driven Hierarchical Reinforcement Learning (DHRL) framework with a high-level L-Agent selecting learning items and a low-level P-Agent choosing associated practice items, incorporating a knowledge state estimation module (DIMKT) to track evolving student knowledge. A communication mechanism between the agents ensures smooth coordination, adapts to item difficulty, and produces efficient, smooth learning paths that maximize learning effectiveness.
        \item \textbf{PKSD.} PKSD introduces a privileged knowledge state distillation framework for reinforcement learning-based educational path recommendation, where the student’s simulator-derived knowledge state is used as privileged information during training. A two-stage process first trains a knowledge state encoder with the privileged state and then trains a GRU-based knowledge state adapter to estimate the latent encoding from regular exercise logs during inference. GEPKSD further incorporates a knowledge graph encoder, combining concept dependencies with learner states, enabling RL agents to generate personalized, structured learning paths that adapt effectively to diverse student knowledge levels.
        \item \textbf{KnowLP.} KnowLP constructs personalized learning paths using a step-by-step reinforcement learning framework guided by dual knowledge structure graphs capturing both prerequisite and similarity relations among knowledge concepts. Three specialized agents—prerequisite, similarity, and difficulty—sequentially select knowledge concepts and corresponding exercises while simulating the discrimination learning process, ensuring that blocked or confusing concepts are addressed. A knowledge tracing module (DIMKT) dynamically tracks student mastery and exercise difficulty, allowing the system to adaptively generate coherent and effective learning paths that improve student progression.
    \end{itemize}

    \begin{table}[htbp]
    \centering
    \caption{Hyper-parameter search space of ILER methods.}
    \label{tab:hyper-parameter-ILER}
    \begin{tabular}{lll}
    \toprule
    Algorithm & Hyper-parameter & Search Space \\
    \midrule
    \multirow{5}{*}{\textbf{General Settings}} 
    & DKT embedding size & 32, 64, 128, 256 \\
    & DKT hidden size & 64, 128, 256, 512 \\
    & DKT layers & 1, 2, 3, 4 \\
    & DKT dropout rate & 1e-5, 1e-4, 1e-3, e-2 \\
    & Episodes & 2000, 5000, 10000\\
    \midrule
    \multirow{6}{*}{\textbf{AKTRec}} 
    & Dropout rate & 0.05, 0.1, 0.2, 0.3 \\
    & Model dimension & 64, 128, 256 \\
    & Feed-forward dimension & 128, 256, 512 \\
    & Number of attention heads & 2,4,8 \\
    & Number of blocks & 1, 2, 4 \\
    & Learning rate & 1e-5, 1e-4, 1e-3 \\
    \midrule
    \multirow{6}{*}{\textbf{SimpleKTRec}} 
    & Embedding size & 32, 64, 128, 256 \\
    & hidden size & 32, 64, 128, 256 \\
    & Learning rate & 1e-5, 1e-4, 1e-3 \\
    & Dropout rate & 0.05, 0.1, 0.3, 0.5 \\
    & Number of blocks & 1, 2, 4 \\
    & Number of attention heads & 4, 8 \\
    \midrule
    \multirow{5}{*}{\textbf{DRE}} 
    & exercise embedding size & 30, 50, 70, 100 \\
    & KC embedding size & 5, 10, 15, 20 \\
    & content embedding size & 50, 100, 150, 200 \\
    & hidden size & 64, 128, 256 \\
    & Batch size & 16, 32, 64 \\
    \midrule
    \multirow{4}{*}{\textbf{MMER}} 
    & $\gamma$ & 0.7, 0.8, 0.85, 0.9 \\
    & Hidden size & 200 \\
    & Learning rate & 1e-4,1e-3,5e-3 \\
    & Fine-tuning ratio & 0.1,0.2,0.3 \\
    \midrule
    \multirow{1}{*}{\textbf{KCPER}}
    & $\delta$ & 0.5, 0.6, 0.7, 0.8 \\
    \midrule
    \multirow{7}{*}{\textbf{MULOER}} 
    & Embedding size & 64, 128, 256 \\
    & Number of attention heads & 2, 4, 6, 8 \\
    & Learning rate & 1e-5, 1e-4, e-3\\
    & Dropout rate & 0.1, 0.2, 0.3 \\
    & $N$ & 20, 50, 100 \\
    & $T$ & 100, 200, 300 \\
    & $d$ & 10, 20, 30 \\
    \midrule
    \multirow{4}{*}{\textbf{KG4Ex}} 
    & $\delta_1$ & 0.5, 0.6, 0.7, 0.8 \\
    & $\delta_2$ & 0.5, 0.6, 0.7, 0.8 \\
    & LSTM hidden size & 64, 128, 256 \\
    & Learning rate & 1e-4, 1e-3, 1e-2 \\
    & Batch size & 32, 64, 128, 256 \\
    \midrule
    \multirow{6}{*}{\textbf{ER-TGA}} 
    & Weak KC threshold & 0.35, 0.4, 0.45, 0.5, 0.55, 0.6 \\
    & Crossover probability & 0.6, 0.65, 0.7, 0.75, 0.8, 0.85, 0.9 \\
    & Mutation probability & 1e-6, 1e-5, 1e-4, 1e-3, 1e-2, 1e-1 \\
    & Population size & 30, 40, 50, 60, 70, 80 \\
    & Tribe size & 6, 12, 24, 36, 48, 60 \\
    & Max generation & 10, 30, 50, 100, 150 \\
    \midrule
    \multirow{5}{*}{\textbf{NR4DER}} 
    & $\beta$ & 0.4, 0.6, 0.8, 1.0 \\
    & $\gamma$ & 0.1, 0.3, 0.5, 0.7, 1.0 \\
    & $\delta$ & 0.5, 0.6, 0.7, 0.8 \\
    & Number of attention heads & 2, 4, 6, 8 \\
    & Batch size & 16, 32, 64, 128 \\
    \bottomrule
    \end{tabular}
    \end{table}

    \begin{table}[htbp]
    \centering
    \caption{Hyper-parameter search space of PLER methods.}
    \label{tab:hyper-parameter-PLER}
    \begin{tabular}{lll}
    \toprule
    Algorithm & Hyper-parameter & Search Space \\
    \midrule
    \multirow{3}{*}{\textbf{AC}} 
    & Learning rate & 1e-5, 1e-4, 1e-3 \\
    & Reward discount factor & 0.85, 0.9, 0.95, 0.98 \\
    & Batch size & 16, 32, 64 \\
    \midrule
    \multirow{3}{*}{\textbf{DQN}} 
    & Learning rate & 1e-5, 1e-4, 1e-3\\
    & Reward discount factor & 0.85, 0.9, 0.95, 0.98 \\
    & Batch size & 16, 32, 64 \\
    \midrule
    \multirow{4}{*}{\textbf{CSEAL}} 
    & LSTM dropout rate & 0.1, 0.2, 0.3, 0.4, 0.5 \\
    & Learning rate & 1e-5, 1e-4, 1e-3 \\
    & Reward discount factor & 0.85, 0.9, 0.95, 0.98\\
    & Batch size & 16, 32, 64 \\
    \midrule
    \multirow{5}{*}{\textbf{RLtutor}} 
    & Learning rate & 1e-5, 1e-4, 1e-3 \\
    & Reward discount factor & 0.85, 0.9, 0.95, 0.98 \\
    & Batch size & 16, 32, 64 \\
    & Clipping parameter & 0.1, 0.2, 0.3 \\
    & Value function coefficient & 0.2, 0.4, 0.5, 0.6, 0.8 \\
    \midrule
    \multirow{4}{*}{\textbf{SRC}} 
    & Embedding size & 32, 64, 128, 256 \\
    & Batch size & 32, 64, 128, 256 \\
    & Dropout rate & 0, 0.1, 0.2, 0.3, 0.4, 0.5 \\
    & Learning rate & 1e-5, 1e-4, 1e-3 \\
    \midrule
    \multirow{3}{*}{\textbf{GEHRL}}
    & Learning rate & 1e-5, 1e-4, 1e-3 \\
    & $\alpha$ & [0, 1.0] \\
    & $\beta$ & [0, 1.0] \\
    \midrule
    \multirow{4}{*}{\textbf{DLPR}} 
    & DIMKT embedding size & 64, 128, 256 \\
    & DIMKT difficulty level & 10, 50, 100 \\
    & DIMKT learning rate & 1e-4, 1e-3, 2e-3 \\
    & Answering correctly probability & [0.4, 0.6] \\
    \midrule
    \multirow{3}{*}{\textbf{PKSD}} 
    & Learning rate & 1e-5, 1e-4, 1e-3 \\
    & Reward discount factor & 0.85, 0.9, 0.95, 0.98 \\
    & Batch size & 16, 32, 64 \\
    \midrule
    \multirow{5}{*}{\textbf{KnowLP}} 
    & DIMKT embedding size & 64, 128, 256 \\
    & DIMKT difficulty level & 10, 50, 100 \\
    & DIMKT learning rate & 1e-4, 1e-3, 2e-3 \\
    & Answering correctly probability & [0.4, 0.6] \\
    & Improvement threshold & 1e-4, 1e-3, 1e-2 \\
    \bottomrule
    \end{tabular}
    \end{table}

\section{Detailed Information of UniER}
\subsection{Additional Experimental Details} \label{Additional Experimental Details}
\noindent\textbf{Implementation Details.} To ensure a comprehensive evaluation and maintain fairness across a broad spectrum of models, we introduce UniER, an open-spource toolkit.
This toolkit is built on top of Pytorch 2.8.0~\cite{PasGM2019}.

\noindent\textbf{Hardware Specifications.} All our experiments were carried out on a Linux server with an Intel(R) Xeon(R) Silver 4410Y CPU, 256GB of RAM, and four NVIDIA GeForce RTX 4090 GPUs (24GB VRAM each).

\noindent\textbf{Hyperparameter Settings.} Tables \ref{tab:hyper-parameter-ILER} and \ref{tab:hyper-parameter-PLER} provide a comprehensive list of all hyperparameters used in our random search complete with their search spaces. For the design of the default hyperparameters please refer to our code base in https://github.com/chanllon/UniER.

This experimental setup enables us to measure the efficiency of each method robustly, providing a clear reflection of their practical viability in real-world applications.

%%%%%%%%%%%%%%%%%%%%%%%%%%%%%%%%%%%%%%%%%%%%%%%%%%%%%%%%%%%%

%\newpage
%\input{checklist.tex}

\end{document}